\documentclass[11pt,eps]{article}
\usepackage{amssymb}
\usepackage{amsmath}
\usepackage{fullpage}
\usepackage{epsfig}
\usepackage{graphicx}
\newtheorem{definition}{Definition}
\newtheorem{theorem}{Theorem}

\newtheorem{lemma}{Lemma}

\newtheorem{corollary}{Corollary}

\title{On Optimal Deadlock Detection Scheduling}
\author{Yibei Ling$^\star$, Shigang Chen$^\diamond$,  Cho-Yu Jason Chiang$^\star$ \\ \\
$^\star$Applied Research Laboratories, Telcordia Technologies, \\ \{lingy,chiang\}@research.telcordia.com \\
$^\diamond$Department of Computer \& Information Science \& Engineering \\
University of Florida \\ 
sgchen@cise.ufl.edu 
}
\date{2006}
\begin {document}
\maketitle
\begin{abstract}
Deadlock detection scheduling is an important, yet often overlooked problem
that can significantly affect the overall performance of deadlock handling.  
Excessive initiation of deadlock detection 
increases overall message usage, 
resulting in degraded system performance in the absence of deadlocks; 
while insufficient initiation of deadlock detection    
increases the deadlock persistence time, resulting in an increased  
deadlock resolution cost in the presence of deadlocks.
The investigation of this performance tradeoff, however, 
is missing in the literature.  
This paper studies the impact of deadlock detection scheduling on 
the overall performance of deadlock handling. In particular, we show 
that there exists an optimal deadlock detection frequency that
yields the minimum long-run mean average cost,
which is determined by
the message complexities of the deadlock detection and resolution 
algorithms being used, 
as well as the rate of deadlock formation, denoted as $\lambda$.   
For the best known deadlock detection and resolution algorithms,  
we show that the asymptotically optimal frequency of deadlock 
detection scheduling that minimizes the overall message overhead is ${\cal O}((\lambda n)^{1/3})$, 
when the total number $n$ of processes is sufficiently large. 
Furthermore, we show that in general  
fully distributed (uncoordinated) deadlock detection scheduling cannot be 
performed as efficiently as centralized (coordinated) deadlock detection scheduling. \\ \\
\end{abstract} \footnote{The material in this paper was presented in part at the  
Twenty-Fourth Annual ACM SIGACT-SIGOPS Symposium on
Principles of Distributed Computing, Las Vegas, Nevada, July 17-20, 2005}

\noindent {\bf Keywords/Index Terms:}
Deadlock detection scheduling, Deadlock formation rate, Deadlock persistence time \\

\addtolength{\baselineskip}{\baselineskip}
\section{Introduction}

The distributed deadlock problem 
\cite{Gray1981,Massey1986,Lee2001,Singhal1989,Krivokapic1999,Kshemkalyani1999}
arises from resource contention introduced 
by concurrent processes in distributed computational environments.
It has received a great deal of attention in
different areas such as distributed computing theory \cite{Obermark1982,Singhal1989,Kim1997}, 
distributed database \cite{Lin1996,Kshemkalyani1999,Gray1981,Knapp1987,Krivokapic1999}, 
and parallel and distributed simulation 
\cite{Boukerche2002,Tropper1993,Misra1986}.
A deadlock is a persistent and circular-wait condition, where
each process involved in a deadlock waits indefinitely for resources held by other processes 
while holding resources 
needed by others. As a result, none of the processes waiting for needed resources can 
continue computation any further without 
obtaining the waited-for resources.  
A deadlock has an adverse performance effect that
offsets the advantages of 
resource sharing and processing concurrency.

There are three common strategies of dealing with the
deadlock problem: {\it deadlock prevention}, {\it deadlock avoidance}, and {\it deadlock
detection and resolution}.  It is a long-held consensus that 
both deadlock
prevention and deadlock avoidance strategies
are conservative and less feasible in handling the deadlock problem 
in general, whereas the deadlock detection/resolution strategy
is widely accepted as an optimistic and 
feasible solution to the deadlock problem, because of its 
exclusion of the unrealistic 
assumption about resource allocation requirements 
of the processes \cite{Knapp1987,Lee2001,Singhal1989,Gonzelez1999,Terekhov1999}. 
The central idea behind the deadlock detection and resolution strategy is that it
does not preclude the possibility of deadlock occurring
but leaves the burden of minimizing the adverse impact of deadlock 
to deadlock detection and resolution mechanisms. 
Under this scheme,  
the presence of deadlocks is detected by a periodic initiation of a deadlock detection 
algorithm and then resolved by a deadlock 
resolution algorithm \cite{Wang1995,Terekhov1999,Gonzelez1999}. 

Despite significant performance improvement in the past, 
deadlock detection remains a costly operation 
\cite{Singhal1989,Krivokapic1999,Macri1976}. It requires
dynamical maintenance of wait-for-graph  (WFG) that reflects the 
runtime wait-for dependency among distributed processes, 
and performs a graph analysis to detect the presence of deadlocks.
There is a substantial tradeoff between the cost of 
deadlock detection and that of deadlock resolution \cite{Singhal1989,Lee2001,Park1992}.
An initiation of deadlock detection consumes 
runtime system and network resources which are basically pure overheads 
when no deadlock is present \cite{Singhal1989,Macri1976}.
Excessive initiation of 
deadlock detection would reduce the
deadlock resolution cost but result in system 
performance degradation in the absence of 
deadlock, while infrequent deadlock detection 
would be accompanied by the increased deadlock size, resulting in an increased deadlock resolution cost in the presence
of deadlocks \cite{Park1992,Lee2001,Lee2004,Baldoni1997}. 
It is evident that {\it deadlock detection scheduling} is one of 
key factors affecting the overall system performance of 
deadlock handling.   
Nevertheless, to the best of our knowledge, this subject 
is generally missing in the literature.

This paper investigates the optimal deadlock detection scheduling. 
We study how to best schedule deadlock detections 
so as to minimize the long-run mean average cost of deadlock handling. 
We formulate this problem by introducing a generic cost model (utility metric) and
use this cost model to  
establish a connection between 
deadlock detection and 
deadlock resolution costs, in relation to the rate of deadlock formation.
We show that there exists a unique 
optimal deadlock detection frequency that yields 
the minimum long-run mean average cost.
Moreover, our result indicates that 
the asymptotically optimal frequency of deadlock 
detection that minimizes the message overhead
is ${\cal O}((\lambda n)^{1/3})$, 
when the number $n$ of processes in the system is sufficiently large.
In addition, we prove that 
a fully distributed (uncoordinated) detection scheduling 
can not be performed as efficiently as its centralized counterpart (coordinate scheduling).
  
The rest of this paper is organized as follows. 
Section 2
contains a brief summary of the distributed 
deadlock detection and resolution algorithms. 
Section 3 gives the notions and definitions. 
Section 4 provides the
detailed mathematical analysis and proves
the existence and uniqueness of an optimal detection frequency. 
The determination of the optimal deadlock
detection frequency, its asymptotic relation with the number of processes in 
a distributed system, and the impact of random detection scheduling upon the long-run 
mean average cost of deadlock handling,  are presented. 
In Section 5,  the main contribution of this paper is highlighted and  
the possible future work is discussed. 

\section{Background}

In this section we provide a brief summary of 
worst-case analysis of existing distributed 
detection algorithms of generalized deadlocks
and deadlock resolution algorithms 
since some results will be used
later on. 
We also touch on Gray's simulation model \cite{Gray1981}
as well as Massey's formulation \cite{Massey1986}.  

We restrict our discussion to distributed detection and resolution algorithms.
The references \cite{Knapp1987,Kshemkalyani1994,Kshemkalyani1997,Krivokapic1999,Kshemkalyani1999,Lee2001} 
provide excellent gateways to the 
state of the art in this area for the generalized resource 
request model. 
In the following, we give a brief summary of the
worst-case performance of the existing distributed 
detection algorithms. 

\begin{table}[htb]
\centering
\begin{tabular}{||c|c|c|c|c||} \hline \hline 
Criterion & Bracha- & Wang \cite{Wang1990} & \multicolumn{2}{c||}
{Kshemkalyani \&  Singhal} \\ 
& Toueg \cite{Bracha1987} & et al.  &  \cite{Kshemkalyani1994}  & 
 \cite{Kshemkalyani1999}\\ \hline \hline 
Phases & $2$ & $2$ & $1$ &  $1$  \\ \hline 
Delay & $4d$ & $3d+1$ & $2d+2$ & $2d$ \\ \hline 
Message & $4e$ & $6e$ & $4e-2n+2l$  & $2e$ \\ \hline  \hline
\end{tabular} 
\caption{Distributed Deadlock Detection Algorithms}
\end{table}

Table~1 summarizes the worst-case complexities 
of distributed 
deadlock detection algorithms \cite{Bracha1987,Wang1990,Kshemkalyani1994,Kshemkalyani1999},
where $n$ is the total number of processes, 
$e$ the number of edges, $d$ the diameter, and $l$ the number of sink nodes of the WFG. 
The distributed detection algorithm for generalized 
deadlocks by Kshemkalyni and Singhal \cite{Kshemkalyani1999}
is the clear winner among the algorithms listed in Table~1. 
Their algorithm 
has achieved a message complexity of $2e$ and a time complexity of $2d$,
which are believed to be optimal. 
Since $e = n (n-1)$ and $d = n$ in the worst-case analysis,
the worst-case message complexity and time complexity thus can be 
written as $2 n^2$ and $2n$, respectively. 

Although deadlock detection and deadlock resolution are 
often discussed separately,    
the latter is as important as the former 
\cite{Knapp1987,Singhal1989,Kshemkalyani1994,Gonzelez1999,Terekhov1999,Warnakulasuriya2000,Lee2001}.
The primary issue of deadlock resolution \cite{Lee2004,Lee2001,Lin1996}
is to selectively abort a subset of 
processes involved in the deadlock so as to  minimize the overall abortion cost 
\cite{Macri1976,Singhal1989,Terekhov1999,Gonzelez1999}.
This is often referred to as the {\it minimum abort set problem}.
These victim (aborted) processes must cancel all 
pending requests and release all the 
acquired resources in order to avoid false deadlock detection 
and resolution \cite{Singhal1989,Kshemkalyani1994,Gonzelez1999}. 
The abortion cost thus includes (1) 
the sending of cancel messages to those resources, 
and (2) the sending of reply messages to all the waiting 
processes that are currently being blocked for the 
resources held by the aborted processes.
One noteworthy point is that these waiting processes could be either 
transitively blocked or deadlocked processes.  
To further reduce the abortion cost,  
checkpointing is sometimes introduced to 
prevent the victim processes 
from being rolled back from
scratch \cite{Ling2001}. 

\noindent In addition, it is possible that more than two processes 
can independently detect the same deadlock. If each process that detects 
a deadlock resolves it, then the deadlock resolution will be highly inefficient 
and will result in subsequent false deadlock detection and deadlock resolution 
\cite{Singhal1989,Gonzelez1999,Kshemkalyani1997,Lee2004}.
Therefore, only one process should be selected for 
resolving a deadlock, which in turn requires that the initiations of deadlock 
resolution algorithm
in different sites be coordinated. 
Such a coordination for safe deadlock resolution 
comes at an additional communication cost in message exchange \cite{Gonzelez1999}. 

\noindent Generally, deadlock resolution cost is measured either 
in terms of time complexity \cite{Ling2005,Lin1996,Terekhov1999},
or in terms of message complexity \cite{Lee2004,Lee2001,Gonzelez1999}.
The complexity of resolution algorithms is summarized in Table~2,
where $n$ is the total number of processes, $m$ the number of processes 
having the priorities greater than deadlocked processes, $N_r$ the number of resources, and 
$n_D$  the size of a deadlock.
Note that the message complexities are not given in 
\cite{Lin1996,Terekhov1999}.

\begin{table}[htb] \label{tab:resolution}
\centering
\begin{tabular}{||c|c|c|c||} \hline \hline 
Complexity & Lin & Terekhov \&    & Mendivil \\  
 &  \& Chen \cite{Lin1996} &  \& Camp \cite{Terekhov1999}  & 
{\it et al.} \cite{Gonzelez1999} \\ \hline \hline 
Time  & ${\cal O}(n_D^3)$ & ${\cal O}(n^3N_r)$ &  ${\cal O}(mn_D)$ \\ \hline 
Message &  &  & ${\cal O}(m n_D^2)$ \\ \hline \hline
\end{tabular}
\caption{Distributed Deadlock Resolution Algorithms} 
\end{table} 

By transforming the problem of deadlock resolution into 
a {\it minimum vertex cut problem},
Lin \& Chen's algorithm \cite{Chen1996}
can identify an optimal set 
of victim processes to be aborted, with the properly selected abortion 
cost to avoid 
the starvation and livelock problems. 
The main feature of Terekhov \& Camp's algorithm is to take  
the number of resources into account. 
The deadlock resolution algorithm proposed by Mendivil 
{\it et al.} \cite{Gonzelez1999} uses
a probe-based approach, with a focus on the safety aspect of deadlock 
resolution. 
The novelty of this algorithm is to use 
an additional round of message exchanges to 
gather the information needed for efficient resolution after deadlocks are detected.
The algorithm uses special message known as probes 
to travel in the opposite direction of  
the edges in AWFG (asynchronous wait-for graph), 
and then chooses the lowest priority 
process of each detected cycle as a victim process to be aborted,
hence avoiding
the livelock and starvation problems.
This deadlock resolution algorithm \cite{Gonzelez1999}
excels in the use of formal methods to prove the
correctness and in its fine-granular analysis of the algorithm complexities.
In particular, its message complexity is of ${\cal O}(mn^2_D)$.
The worst-case message complexity can also be written as
${\cal O}(n^3)$ because the eventual deadlock size, $n_D$, is bounded
by the total number of processes in the distributed system, that is, 
$m = {\cal O}(n)$ and $n_D = {\cal O}(n)$.

The past research has been primarily aimed at 
minimizing the complexities (costs) of the deadlock detection and resolution algorithms.
Although deadlock detection scheduling (particularly how frequently deadlock detection
should be performed) 
has significant impact on the overall performance of deadlock handling in practice, 
it is not explicitly studied but rather implicitly reflected 
in the description of deadlock detection algorithms, without a clear guideline. 
For instance, in 
\cite{Knapp1987,Singhal1989,Kshemkalyani1999,Lee2001,Macri1976,Chandy1983,Knapp1987,Chen1996}, the authors stated that
a deadlock detection is initiated when a deadlock is suspected. 
Other works \cite{Park1992,Krivokapic1999} suggested that
it would be highly inefficient if 
deadlock detection is performed whenever 
a process/transaction becomes blocked.

The performance of deadlock handling not only depends on 
the per-detection cost of the deadlock detection algorithm, 
but also on how frequently the deadlock detection algorithm is executed 
\cite{Krivokapic1999,Park1992,Macri1976}. 
The choice of deadlock detection frequency presents a 
tradeoff between deadlock detection cost and deadlock resolution
cost \cite{Knapp1987,Singhal1989,Park1992,Lee2001,Krivokapic1999}. 
Park {\it et al.} \cite{Park1992} 
pointed out that the reduction of deadlock resolution cost can be achieved 
at the expense of deadlock detection cost. 
Krivokapie {\it et al.} \cite{Krivokapic1999} showed in their simulation study that 
the path-pushing algorithm (one type of deadlock detection algorithm) 
is highly sensitive to the frequency of deadlock detection.
Gray {\it et al.} \cite{Gray1981} showed that the probability of a transaction waiting for a lock
request is rare. They used a ``straw-man analysis" in 
their simulation 
model that agreed well with the observation on several 
data management systems. 
Massey \cite{Massey1986} formulated a probabilistic model 
that gave an analytic justification for the simulation results
reported in \cite{Gray1981}, 
showing that the probability of deadlock grows
linearly with respect to the number of transactions 
and grows in the fourth power of the average number of resources required
by transactions. 
 
To our best knowledge, only a few papers 
\cite{Gray1981,Lee2001,Terekhov1999,Chen1996,Singhal1989,Macri1976,Ling2005} mentioned about
deadlock detection scheduling but under a different context from this paper.
The idea of relating deadlock recovery cost to deadlock persistence time, 
and identifying an optimal deadlock detection frequency that minimizes 
the long-run mean average cost from the perspective of deadlock handling, has not been considered 
before. 

\section{Deadlock Persistence time and Deadlock Recovery Cost} 

In this section, 
we first give the following definitions in order to simplify problem formulation.

\begin{definition} \label{def:independ}
A deadlock refers to a circular-wait condition where a set of processes
waits indefinitely for resource from each other.
A blocked process (a process in a deadlock) refers to the process 
that waits indefinitely on other processes to progress. 
Deadlock size refers to the total number of blocked processes involved in 
the deadlock. 
\end{definition}

\noindent Blocked processes can be decomposed into two categories: {\it deadlocked} and
{\it transitively blocked} processes \cite{Lee2001}.
Deadlocked processes belong to a cycle in the WFG, while 
a transitively blocked process refers to one that waits for the 
resources held by other processes but does not belong to any cycle in the WFG.  	

\begin{definition}
Two deadlocks are said to be independent of each other if they don't share any
deadlocked process.   
\end{definition} 

The independence of deadlock occurrence can be
justified by 
the wide acceptance of large-scale distributed systems 
and adoption of fine-granularity locking mechanism such as  
{\it semantic locking} \cite{Roesler1988,Krivokapic1999} and 
record-granularity locking \cite{Roesler1988}.
After decades of research and development,
large-scale distributed systems 
allow resource sharing among 
hundreds or even thousands of sites across a network \cite{Roesler1988,Krivokapic1999}.
The fine-granular locking mechanisms enable a higher
degree of parallelism. 
Large-scale resource distribution and fine-granularity of locking
make deadlocks likely to form independently.

\noindent Now we are in a position to introduce 
the notion of deadlock persistence time which serves as a basis 
for our problem formulation. Let $S=\{S_1, S_2,\cdots \}$ be the time instants at
which independent deadlocks initially occur, i.e., the $i$th deadlock forms at time $S_i$.  

\begin{definition} \label{def:definition1}
The persistence time of the $i$th deadlock with respect to time $t$, denoted by $t_p(t,S_i)$, 
is 
\[
t_p(t,S_i) = 
\left 
\{ 
\begin{array}{ll}
t-S_i, & t > S_i; \\     
0, & t \leq S_i 
\end{array}  \right.
\]  
\end{definition} 
The function $t_p(t,S_i)$ represents the time interval between the present time and the 
time at which  the deadlock is initially formed. It grows linearly 
until the deadlock is resolved. The notion of deadlock persistence time 
in spirit is similar to that of {\it deadlock latency} or 
{\it deadlock duration} in \cite{Lee2001,Lee2004}.
  
Once a deadlock is formed,   
other processes requesting resources currently held by the blocked processes
in the deadlock (including deadlocked and transitively blocked processes)  
will be blocked forever unless the deadlock is resolved. 
As a result, each deadlock acts as an attractor to trap 
more processes into it.
As the deadlock persistence time increases, 
the size of the deadlock (the total number of processes 
involved in the deadlock)
keeps growing \cite{Singhal1989,Kim1997,Lee2001,Lee2004}, 
which in turn increases the deadlock resolution cost. 
\begin{figure}[th]
\centerline{\epsfig{file=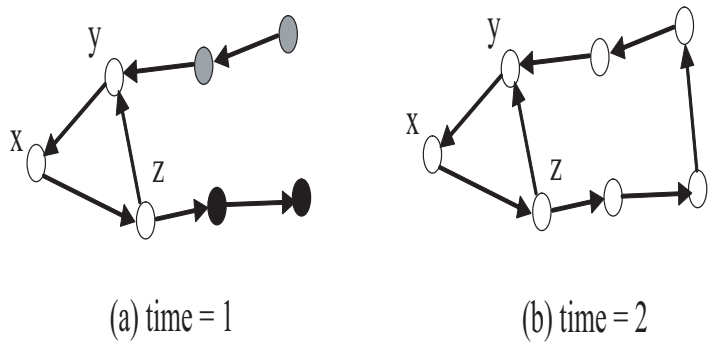,height=2.5in,width=5in}}
\caption{Increasing Deadlock Size with Deadlock Persistence Time}
\label{fig:fig0}
\end{figure}

\noindent This dependency of deadlock resolution cost 
upon deadlock persistence time can be illustrated in the example in 
Fig(\ref{fig:fig0}). At time=1, there 
are three circularly deadlocked processes and 
two transitively blocked processes. 
At time=2, there are seven circularly deadlocked processes. 
The graphs (a) and (b) in Fig(\ref{fig:fig0}) represent two snapshots 
in the wait-for graph, showing that the deadlock size (including both deadlocked and 
transitively-blocked processes) grows with the deadlock persistence time.
Intuitively, a deadlock resolution algorithm will 
have to explore the entire deadlock  
in order to identify the least costly set of victim processes to be 
aborted. 
The intrinsic dependency of deadlock size 
(and thus deadlock resolution cost) upon deadlock 
persistence time was observed by Singhal {\it et al.} 
\cite{Singhal1989,Kshemkalyani1997,Villadangos2003}, Lee \cite{Lee2001,Lee2004},
Krivokapic {\it et al.} \cite{Krivokapic1999}, 
Lin {\it et al.} \cite{Lin1996}, and Park {\it et al.} \cite{Park1992}.

Throughout this paper, we  
use $n$ to denote the total 
number of processes in a distributed system and 
$n_D(.)$ to denote the size of a deadlock.
Consider an arbitrary deadlock. Its size is 
a function of deadlock persistence time $t_p$, denoted as $n_D(t_p)$.  
The deadlock size $n_D(t_p)$ by nature is a discrete staircase function 
that jumps by one whenever a new process becomes transitively blocked
by the deadlocked processes. To facilitate our mathematical analysis, 
we will treat $n_D(t_p)$ instead as a continuous, increasing function, which is 
an approximation of the staircase one. 

The deadlock size function $n_D(t_p)$ has the following mathematical properties.
(1) $n_D(0)=0$, (2) monotonicity: $n_D^\prime(t_p) > 0, \ t_p \geq 0$,
and (3) bounded: $n_D(\infty) \leq n$, 
where $n^{\prime}_D(t_p)$ is the derivative of $n_D(t_p)$.
The first property refers to the initial deadlock size
at $t_p=0$ is zero. The second property 
reflects the fact that the number of blocked processes in the deadlock increases monotonically 
with deadlock persistence time $t_p$, and 
the third property indicates that the eventual deadlock size is bounded by 
the total number of distributed processes.  For the sake of easy presentation, 
we drop the subscript $p$ hereafter. 

Now let's revisit the message complexity achieved by 
the deadlock resolution algorithm proposed by Mendivil {\it et al.} 
\cite{Gonzelez1999}, which is ${\cal O}(mn_D^2) = {\cal O}(n n_D^2)$,  
where $m$ is the number of deadlocked processes having priority values 
greater than those of the deadlocked processes. 
Notice that the deadlock size, $n_D$, is
a function of deadlock persistence time. 
To make this dependency concrete, the message overhead 
can be written as $c n n^2_D(t)$ for some constant $c$.
This result will be used later to derive the optimal frequency of deadlock detection scheduling. 

\section{Mathematical Formulation}

In this section, we begin with a generic cost model that accounts for
both deadlock detection and deadlock resolution, 
which is independent of deadlock detection/resolution
algorithms being used.  We then prove the existence and the uniqueness of an optimal 
deadlock detection frequency that  
minimizes the long-run mean average cost 
in terms of the message complexities of the best known deadlock detection/resolution algorithms.  

In this paper we choose the message complexity as the 
performance metric for 
measuring the detection/resolution cost. 
The reason for choosing message complexity is that communication overhead is 
generally a dominant factor that affects the overall system performance 
in a distributed system \cite{Singhal1989, Knapp1987,Kshemkalyani1997,Kshemkalyani1999},
as compared with processing speed and storage space.   
Note that the worst-case message complexity can normally be expressed 
as a polynomial of $n$. 
Per deadlock detection cost is denoted as $C_D$.
The resolution cost for a deadlock is denoted as $C_R(t)$, which 
is a function of the deadlock persistence time $t$. 
In general, the resolution cost is a polynomial of $n_D(t)$. 
For example, the deadlock resolution cost for 
Mendivil's algorithm \cite{Gonzelez1999}
is $c n n^2_D(t)$. 
Because $n_D(t)$ is a monotonically increasing function of deadlock 
persistence time. $C_R(t)$ is also monotonically increasing with deadlock persistence time. 
We assume that deadlock formation follows a Poisson process for two reasons:
First, the Poisson process is widely used to approximate
a sequence of events that occur randomly and independently. 
Second, it is due to mathematical tractability of the Poisson process, which allows
us to characterize the essential aspects of complicated processes while
making the problem analytically tractable. 

The following theorem presents the long-run mean average cost of deadlock handling
in connection with the rate of deadlock formation and the frequency of deadlock detection. 
\begin{theorem} \label{theo:theorem1}
Suppose deadlock formation follows a Poisson process with
rate $\lambda$. The long-run mean average cost 
of deadlock handling, 
denoted by $C(T)$, is  
\begin{align}\label{eq:first}
C(T) = \frac{C_D}{T} + \frac{\lambda \int_0^T C_R(t) dt} {T},
\end{align} 
where the frequency of deadlock detection scheduling is $1/T$. \hfill $\blacktriangle$ 
\end{theorem} 
\noindent Proof:   Let $\{X_i, i\geq 1\}$ be the interarrival times of 
independent deadlock formations, where
random variables $X_i, i\geq 1$ are independent and exponentially 
distributed with mean $1/\lambda$.
Define $S_0=0$ and $S_n=\sum\limits_{i=1}^n X_i$, where
$S_n$ represents the time instant at which the {\it n}th independent deadlock occurs. 

Let $N(t)=\sup \{n:~S_n\leq t \}$ represent
the number of deadlock occurrences within the time interval $(0,t]$. 
The long-run mean average cost is  
\begin{equation}\label{equ:cost1}
\lim\limits_{t\to\infty} \frac{E 
(\hbox{random~cost~in~}(0,t])}{t}, 
\end{equation}
where $E$ is the expectation function. 
In order to associate this cost with the deadlock detection frequency ($1/T$),
we partition the time interval $(0, t]$ into 
non-overlapping subintervals of length $T$.
Let $\xi_k(T)$ be the cost of deadlock handling on the 
subinterval $((k-1)T, kT], \ k > 0$. $\xi_k(T)$ is a random variable.
According to the stationary and independent increments of 
Poisson process \cite{Ross1996},
$E(\xi_i(T)) = E(\xi_j (T)), \ i \neq j$.  
The long-run mean average cost becomes 
\begin{align}\label{equ:equcost}
 C(T) & =
\lim\limits_{t\to\infty} \frac{E 
(\hbox{random~cost~in~}(0,t])}{t} =\lim\limits_{t\to\infty} 
\frac{E(\sum\limits_{k=0}^{\lfloor \frac{t}{T} \rfloor} 
\xi_k(T))}{t}  \nonumber \\
& =\lim\limits_{t\to\infty} \frac{E(\lfloor \frac{t}{T} \rfloor
\xi_1(T))}{t} = \frac{E(\xi_1(T))}{T},
\end{align}
where $\lfloor x \rfloor$ is the floor function in $x$. 

The cost $\xi(T)$ on interval $(0,T]$  
is the sum of a deadlock detection cost $C_D$ and a deadlock resolution cost for
those deadlocks independently formed within the interval $(0,T]$.
For the {\it i}th deadlock formed at time $S_i \leq T $, 
the resolution cost $C_R(T-S_i)$ is a function of the deadlock persistence time $T-S_i$.
Hence, the accrued total cost over $(0,T]$ 
is  
\begin{align} \label{equ:equlimit1}
\xi (T) 
= C_D + \sum\limits^{N(T)}_{i=1}  C_R(T-S_i) I_{\{N(T)>0\}},
\end{align}
where  $I_{\theta}$ is the indicator 
function whose value is 1 (or 0) if predicate $\theta$ is true (or false).   
Among that, the deadlock resolution cost on interval $(0,T]$ is   
 \begin{align} \label{equ:equim}
\sum \limits^{N(T)}_{i=1} C_R(T-S_i)I_{\{N(T)>0\}} 
= \sum\limits^\infty _{i=1} C_R(T-S_i) I_{\{S_i \leq T\}} 
\end{align}
\begin{align} \label{equ:equi}
E\left ( C_R(T-S_i)I_{\{S_i\leq T\}}\right ) = \int\limits^T_0 C_R(T-t)f_i(t)dt 
\end{align}
where $f_i(t)$ is the probability density function of $S_i$ 
which follows
the gamma distribution given below:
\begin{equation}\label{equ:gamma}
f_i(t) = \frac{ \lambda^i}{(i-1)!} t^{i-1}e^{-\lambda t},~t>0.
\end{equation}
Substituting Eq(\ref{equ:gamma}) into Eq(\ref{equ:equi}) gives rise to
\begin{align} \label{equ:equ12}
 E \left( C_R (T-S_i) I_{\{S_i\leq T\}}\right ) =  
\int ^T_0 C_R (T-t) \frac {\lambda ^i}{(i-1)!} t^{i-1} e^{-\lambda t}dt.
\end{align}
The expected total resolution cost over the time interval $(0,T]$
is 
\begin{align} \label{equ:eq100}
& E (\sum\limits ^{N(T)}_{i=1} C_R(T-S_i)  I_{\{N(T)>0\}}) 
 = \sum\limits ^\infty _{i=1} \int ^T_0 C_R(T-t) \frac {\lambda ^i t^{i-1}}
{(i-1)!} e^{-\lambda t}dt  \nonumber \\ 
&=\int^T_0 C_R(T-t) \lambda e^{-\lambda t} \left (\sum \limits ^\infty_{i=1} 
\frac {(\lambda t)^{i-1}}{(i-1)!}\right ) dt  
= \lambda \int ^T_0 C_R(T-t) dt = \lambda \int ^T_0 C_R(t) dt.
\end{align} 
Combining Eqs(\ref{equ:equcost}), (\ref{equ:equlimit1}), 
and (\ref{equ:eq100}) yields
\begin{align}\label{equ:equppp}
C(T)=\frac{E(\xi_1 (T))}{T}=\frac{C_D}{T} + \frac { \lambda \int ^T_0 C_R(T-t)dt}{T} 
= \frac{C_D}{T} + \frac {\lambda \int^T_0 C_R(t)dt}{T}.  
\end{align}  
Theorem~\ref {theo:theorem1} is thus established. \hfill  $\blacksquare$ 

Theorem~\ref{theo:theorem1} is mainly concerned with the impact of
deadlock detection  frequency and  deadlock formation rate  
on the long-run mean average cost of overall deadlock handling. 
It is independent of the choice of deadlock detection/resolution algorithms. 
The following corollary is an immediate consequence of  
Theorem~\ref{theo:theorem1}.
\begin{corollary}\label{co:co1}
The long-run mean average cost of deadlock handling is proportional to 
the rate of deadlock formation $\lambda$.   \hfill $\blacktriangle$ 
\end{corollary}
Proof: the proof is straightforward and thus omitted. \hfill $\blacksquare$ 

Theorem~\ref{theo:theorem1} and Corollary~{\ref{co:co1} 
state that the overall cost of deadlock handling is closely associated 
not only with per-deadlock detection cost, and aggregated resolution cost,
but also with the rate of deadlock formation, $\lambda$.
In the following lemma, we will show the existence and uniqueness 
of asymptotic optimal frequency 
of deadlock detection when deadlock resolution is more 
expensive than a deadlock detection in terms 
of message complexity. 

\begin{lemma}\label{lemma:lemma2}
Suppose that the message complexity of deadlock detection 
is ${\cal O}(n^\alpha)$, and that of  deadlock resolution is 
${\cal O}(n^\beta)$. 
If  $\alpha < \beta$, there exists 
a unique deadlock detection frequency $1/T^*$  
that yields the minimum long-run mean average cost when  $n$ 
is sufficiently large.  \hfill $\blacktriangle$ 
\end{lemma}
Proof: Differentiating Eq(\ref{eq:first}) yields
\begin{equation}
C^\prime(T) = - \frac{C_D}{T^2} + 
\frac{\lambda C_R(T)}{T} - \frac{\lambda \int^T_0 C_R(t)dt}{T^2}.
\end{equation}
Define a function $\varphi(T)$ as follows 
\begin{align} \label{equ:varphi}
\varphi (T)  \equiv T^2 C^\prime (T) = -C_D+\lambda TC_R(T)-\lambda \int^T_0 C_R(t)dt. 
\end{align}
Notice that  $C^\prime(T)$ and $\varphi(T)$ share the same sign. Differentiating $\varphi (T)$, 
we have
\begin{align}
\varphi ^\prime (T) = \lambda TC_R^\prime(T)
\end{align}

Because $C_R(T)$ is a monotonically increasing function, 
$C_R^\prime(T) > 0$, which means $\varphi ^\prime (T) > 0$.
Therefore, $\varphi ^\prime (T)$ is also a monotonically increasing function. 
$C_R(T) - C_R(t) \geq 0$ holds iff $T \geq t$.
For any given $0 < \xi < T$,  it has
\begin{align} \label{equ:qwe}
T C_R(T) - \int^T_0 C_R(t) dt & = \int^T_0 (C_R(T)- C_R(t)) dt 
> \int^{\xi}_0 (C_R(T)- C_R(t)) dt \nonumber \\
&  > \int^{\xi}_0 (C_R(T)- C_R(\xi)) dt 
= \xi(C_R(T)- C_R(\xi)).
\end{align}
Applying Eq(\ref{equ:qwe}) to Eq(\ref{equ:varphi}), we have
\begin{align} \label{equ:qqqq}
\varphi(T) & =  -C_D + \lambda (TC_R(T)-\int^T_0 C_R(t)dt) >  -C_D +\lambda \xi (C_R(T)- C_R(\xi)) 
\end{align}
We further have
\begin{align} \label{equ:qqqqq}
\varphi(T) & > -C_D +\lambda \xi C_R(T)(1- \frac{C_R(\xi)}{C_R(T)})  = -C_D +\lambda \xi C_R(T) \theta
\end{align}
where $\theta = (1- C_R(\xi)/C_R(T))$  and $ 0 < \theta < 1$ since 
$C_R(T)$ is monotonically increasing. 
Substituting $C_D = c_1 n^{\alpha}$ and 
$C_R(\infty)  = c_2 n^{\beta}$ in Eq(\ref{equ:qqqqq}), we obtain  
\begin{align}
& \lim\limits_{T \to\infty}\varphi(T)  >  -c_1 n^{\alpha} +
\lambda \xi \theta c_2 n^{\beta} 
\end{align}
Since $\alpha < \beta$, 
$\lim\limits_{T \to \infty}\varphi(T)$ is asymptotically dominated 
by the term $\lambda \xi \theta c_2 n^{\beta}$ when  $n$ 
is sufficiently large.  Observe that $\varphi(0) = -C_D < 0$, 
and $\varphi(T)$ is monotonically increasing. 
By the intermediate value theorem, it must be true that 
there exists a unique $T^*$, $0<T^*<\infty$,  
such that 
\[ \varphi(T) = T^2 C^\prime(T)= \left \{ \begin{array}{ll}
<0, & 0\leq T<T^* \\
 = 0, & T=T^* \\
>0, & T > T^* .
\end{array}
\right . \]
\noindent
It means that $C(T)$ reaches its minimum at and only at $T = T^*$.
The existence and the uniqueness of
optimal deadlock detection interval  
$T^*=\arg \left ( \min\limits _{T>0} C(T)\right )$ is proved.  \hfill $\blacksquare$

To make the idea behind this derivation concrete,  we apply 
the up-to-date results of deadlock detection/resolution algorithms.
As discussed before, the best-known message complexity of 
a distributed deadlock detection algorithm is $2n^2$ \cite{Kshemkalyani1999} 
when it is written as a polynomial of $n$. 
The best-known message complexity of a deadlock resolution algorithm is   
${\cal O}(nn_D^2)$ \cite{Gonzelez1999}.
Therefore, $C_D = n^2$, and
$C_R(t) = c n n^2_D(t)$, where  $c$ is a positive constant. 
Because the deadlock size $n_D(t)$ is always bounded by $n$,
from (\ref{equ:qqqq}) we have 
\begin{align} \label{equ:equ99}
\varphi (\infty ) = \lim\limits_{T\to\infty} \varphi (T)
 > -C_D  +\lambda \xi (C_R(\infty)- C_R(\xi)) 
\approx
-2 n^2 + \lambda c \xi n^3.
\end{align}
Note that $\xi$ is a fixed value that can be arbitrarily chosen.
For a sufficiently large $n$, 
Eq(\ref{equ:equ99}) becomes 
\begin{align} \label{equ:eqoo}
\varphi(\infty) \approx \lambda c \xi  n^3  >0 &  
\end{align}
$\varphi (0) = -C_D = - 2 n^2$. Because $\varphi(T)$ is monotonically increasing,
there exists an optimal deadlock detection frequency $1/T^*$ such that
$\varphi(T^*)$ and thus $C^\prime(T^*)$ are zero, which minimizes
the long-run mean average cost $C(T)$ for deadlock handling. 

The motivation behind the proof  
is that the cost per deadlock detection is fixed when the total number of processes 
in the distributed system is given, 
while the cost of deadlock resolution monotonically increases with deadlock persistence time.
The resolution cost will eventually outgrow the detection cost if deadlocks persist. 
As we set the time interval $T$ between any two consecutive detections longer, the detection
cost becomes smaller due to less frequent executions of the detection algorithm, 
but the resolution cost becomes larger due to the growth in deadlock size.
This implies that there exists a unique deadlock detection frequency $1/T^*$ that 
balances the two costs such that their sum is minimized.  
The condition that the asymptotic deadlock resolution cost, $C_R(\infty)$, is greater than the cost of deadlock detection, $C_D$,  
constitutes the natural mathematical basis to justify 
distributed deadlock detection algorithms. 

We are now ready to state the asymptotically optimal frequency for deadlock detection based on 
the up-to-date results of distributed deadlock detection and resolution algorithms. 
Recall that the best-known message complexity for 
distributed deadlock detection algorithms is $2 n^2$ \cite{Kshemkalyani1999} and that for 
deadlock resolution algorithms of  ${\cal O}(n n_D^2)$ \cite{Gonzelez1999}. 

\begin{theorem}  \label{theo:theorem10}
Suppose the message complexity for distributed deadlock detection is $2 n^2$, and that for distributed deadlock resolution is ${\cal O}(nn^2_D(t))$.  
Then the asymptotically optimal frequency for scheduling deadlock detections is 
${\cal O}((\lambda n)^{1/3})$. \hfill $\blacktriangle$ 
\end{theorem}
\noindent Proof:  Assume that 
the deadlock size function $n_D(t)$ is both differentiable and integrable.\footnote{
Recall that $n_D(t)$ is a continuous approximation function whose curves between
``jumping points" can be chosen.} 
Then $n_D(t)$ can be expressed in the form of Maclaurin series as follows:
\begin{align} \label{equ:eqpp}
n_D(t) = 
\sum_{i=0}^{\infty} \frac{n^{(i)}_D(0) t^i}{i!} = \sum_{i=0}^{\infty} c_i t^i,
\end{align}
where $n^{(i)}_D(0)$ denote the {\it i}th derivative of 
the deadlock size function $n_D(t)$ at point zero and $c_i = n^{(i)}_D(0)/i!$. 

By the properties of the deadlock size function $n_D(t)$, we have
$n_D(0)=0$ and $n^\prime_D(0)>0$. It can be easily verified that $c_0= 0$ and 
$c_1 = n^{\prime}_D(0) > 0$.
The resolution cost $C_R(t)$ can be written as $c n n^2_D(t)$ for some constant $c$.
By Theorem~\ref{theo:theorem1}, the long-run mean average cost becomes
\begin{equation} \label{equ:ppp1}
C(T) = \frac{2 n^2}{T}+ \lambda c n \frac{\int_0^T n^2_D(t) dt}{T}.
\end{equation}
Inserting Eq(\ref{equ:eqpp}) into Eq(\ref{equ:ppp1}), we have 
\begin{align} \label{equ:ppp2}
C(T)  =  &\frac{2 n^2}{T}+ \lambda c n^3 T^{-1} 
\int_0^T ( \sum_{i=1}^\infty c_i t^i)^2 dt =   \frac{2 n^2}{T}+ 
\frac{\lambda c n^3 \int_0^T (c_1 t + \sum_{i=2}^\infty c_i t^i)^2 dt}{T}.
\end{align}
Through a lengthy calculation, Eq(\ref{equ:ppp2}) can be simplified as 
\begin{align} \label{equ:ppp3}
C(T) & =\frac{2 n^2}{T}+ c \lambda n^3 
(\frac{c_1^2 T^2}{3} +\frac{2c_1 c_2 T^3}{4}) 
+ c \lambda n^3  (\sum_{i=2}^{\infty}\sum_{j=2}^{\infty} 
\frac{c_ic_j T^{i+j}}{i+j+1}).
\end{align}
Taking derivative of Eq(\ref{equ:ppp3}) with respect to $T$, we have
\begin{align} \label{equ:ppp4}
C^\prime(T) & = -\frac{2 n^2}{T^2}+ c \lambda n^3 (c^2_1\frac{2T}{3} +
\frac{3c_1 c_2 T^2}{2})  +c \lambda n^3 ( \sum_{i=2}^{\infty}\sum_{j=2}^{\infty} 
\frac{c_ic_j(i+j) T^{i+j-1}}{i+j+1}).
\end{align}

By lemma~\ref{lemma:lemma2}, there exists a
unique optimal detection frequency $1/T^*$ when $n$ is 
sufficiently large, such that $C(T^*) \leq C(T), \ \ T \in (0, \infty)$.
We know that $C^\prime(T^*)=0$. Based on (\ref{equ:ppp4}),
we transform $C^\prime(T^*)=0$ to the following equation.
\begin{align} \label{equ:final1}   
\frac{1}{n}  = 
\frac{c \lambda}{2} (\frac{2c^2_1 (T^*)^3}{3} +
\frac{3c_1 c_2 (T^*)^4}{2}  
+ \sum_{i=2}^{\infty}\sum_{j=2}^{\infty} \frac{c_ic_j(i+j) (T^*)^{i+j+1}}{i+j+1}).
\end{align}
Only $n$, $T^*$, and $\lambda$ are free variables and the rest are constants.
By performing the Big-O reduction we obtain 
\begin{align} \label{equ:final2}   
\frac{1}{n} & = \Theta( \lambda ((T^*)^3 + (T^*)^4 + (T^*)^5 + ...))  
\end{align}
When $n$ is sufficiently large and $T^*$ is sufficiently small, we have
\begin{align} \label{equ:final3}   
& \frac{1}{n}  = \Theta( \lambda \frac{(T^*)^3}{1 - T^*}) = {\cal O}(\lambda (T^*)^3) \nonumber \\
& T^*  = \Omega(\frac{1}{(\lambda n)^{1/3}})   
\end{align}

Therefore, the asymptotic optimal deadlock detection frequency $1/T^*$ 
is ${\cal O}((\lambda n)^{1/3})$. \hfill $\blacksquare$ 

\begin{figure}[th]
\centerline{\epsfig{file=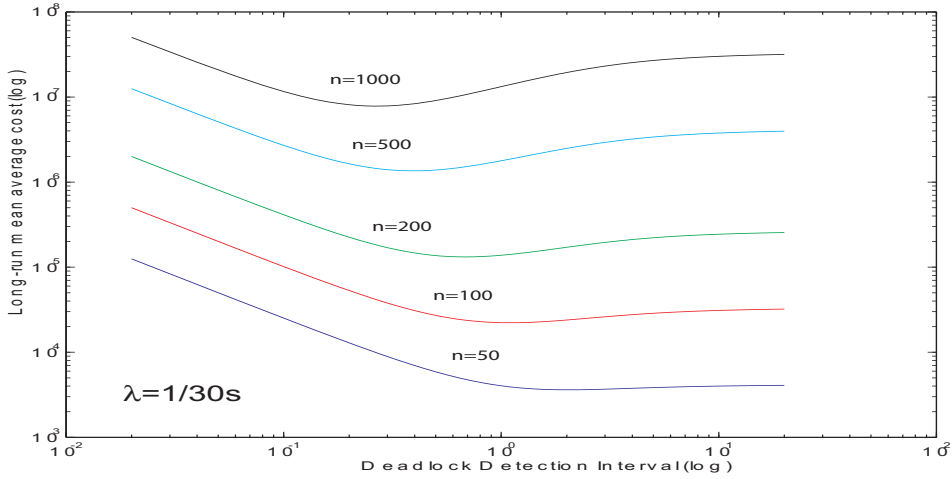,height=2.5in,width=5in}}
\caption{Cost of Deadlock Handling vs. Detection Interval ($n$: number of processes)}
\label{fig:fig1}
\end{figure}
\begin{figure}[th]
\centerline{\epsfig{file=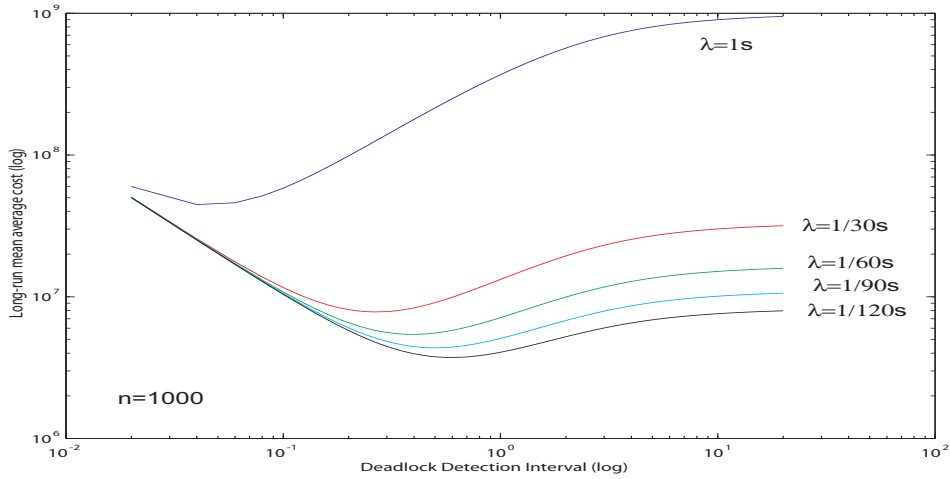,height=2.5in,width=5in}}
\caption{Cost of Deadlock Handling vs. Deadlock Formation Rate $\lambda$}
\label{fig:fig2}
\end{figure}

As an illustration, we consider an example as follows. Let
$C_R(t) = n^3(1-\exp(-t))$, $C_D=n^2$. In accordance with 
Theorem~1,  the long-run mean average cost of deadlock handling thus is  
written as 
\begin{align}
C(T)= \frac{n^2+\lambda n^3(T +\exp(-T)-1)}{T}.
\end{align}

\noindent Figs(\ref{fig:fig1})-(\ref{fig:fig2}) show
log-log plots of a family of curves 
illustrating the dependence of  
long-run mean average cost of deadlock handling upon detection interval.
The {\it x}-axis denotes the deadlock detection interval and the 
{\it y}-axis denotes the long-run mean average cost of deadlock handling.  

\begin{center}
\begin{tabular}{||c|c||} \hline
$\#$ of Processes & Optimal Detection Interval ($\lambda=1$)\\ \hline \hline
$50$ &  $0.214699$(s) \\ \hline
$100$ & $0.148555$(s) \\ \hline
$200$ & $0.103495$(s) \\ \hline
$500$ & $0.064189$(s) \\ \hline
$1000$ & $0.045402$(s) \\ \hline \hline
$\#$ of Processes & Optimal Detection Interval ($\lambda=1/30$)\\ \hline \hline
$50$ &  $2.0223$(s) \\ \hline
$100$ & $1.0973$(s) \\ \hline
$200$ & $0.6832$(s) \\ \hline
$500$ & $0.3942$(s) \\ \hline
$1000$ & $0.2675$(s) \\ \hline \hline
\multicolumn{2}{c}{Table 1: Optimal Detection Interval vs. $\#$ of Processes} \\
\end{tabular}
\end{center}

\noindent In Fig(\ref{fig:fig1}), we present plots of  
the deadlock detection interval and cost of 
deadlock handling under different the total number
of processes, $50,100, 200, 500$, and $1000$, respectively. 
Fig(\ref{fig:fig2}) shows the relationship 
between the overall cost of deadlock handling and 
deadlock detection interval under the different deadlock formation rates,
$1s, 1/30s, 1/60s, 1/90s$, and $1/120s$, respectively. 
Figs(\ref{fig:fig1})-(\ref{fig:fig2}) visualizes convexity that suggests 
the existence of an 
optimal detection frequency, illustrating that 
the overall cost of deadlock handling increases with the total number of 
processes and deadlock formation rate. 

A detailed calculation given in {\it Table~1} shows
that as the number of processes in a distributed system 
increases, the optimal 
detection interval decreases, which is clearly in line with our theoretical analysis. 
In the sequel, we study the impact of coordinated vs. random deadlock
detection scheduling on the performance of deadlock handling. 
We consider two strategies of deadlock detection scheduling: 
(1) centralized, coordinated deadlock detection scheduling, and (2) 
fully distributed, uncoordinated deadlock detection scheduling. 

The centralized scheduling excels in its simplicity in implementation and 
system maintenance, but undermines the reliability and resilience against failures
because one and only one process is elected 
as the initiator of deadlock detections in a distributed system.
In contrast, the fully distributed scheduling excels in 
the reliability and resilience against failures because
every process in the distributed system can  
independently initiate detections \cite{Lee2004}, without a single point
of failure. However, due to the lack of coordination in 
deadlock detection initiation among processes, it presents a 
different mathematical problem from the centralized deadlock 
detection scheduling. 

In the previous discussions we have focused on the derivation of  
optimal frequency of deadlock detection in connection with the rate of
deadlock formation and the message complexities of deadlock detection and 
resolution algorithms, assuming deadlock detections are centrally scheduled 
at a fixed rate of $1/T$. 
To capture the lack of coordination in fully distributed scheduling, 
we will study the case where processes randomly, independently initiate
the detection of deadlocks.   

Let $n$ be the number of processes in a distributed system and 
$T$ be the optimal time interval between any two consecutive deadlock detections
in the centralized scheduling.  
Consider a fully distributed deadlock detection scheduling, where
each process initiates deadlock detection at a rate of $1/(nT)$ independently.  
Although the average interval between deadlock detections in the fully distributed 
scheduling remains $T$ (the same as its centralized counterpart),  
the actual occurring times of those detections are
likely to be non-uniformly spaced because the initiation of 
deadlock detection is performed by
the processes in a completely uncoordinated fashion.  

In the following we will study the fully distributed (random) scheduling 
and compare it with the centralized scheduling. 
Consider a sequence of independently and identically distributed 
{\it iid} random variables $\{Y_i, i\geq 1\}$ defined on $(0,\infty)$ following certain distribution $H$.
The sequence $\{Y_i, i\geq 1\}$ represents the inter-arrival times of deadlock detections
initiated by the fully distributed scheduling, and
it is assumed to be independent of the arrival of deadlock formations.
It is obvious that the centralized scheduling is a 
special case of the fully distributed scheduling.  

Let ${\cal H}$ be the family of all distribution functions on $(0,\infty)$ with 
finite first moment. Namely,  
\begin{align}
&{\cal H}=\left \{H\colon H \hbox{~is~a~CDF~on~$(0,\infty)$,}
\int^\infty _0\bar{H}(t)dt <\infty \right \}  \\
&\hbox{where $\bar{H}(t)\equiv 1-H(t),~\forall t\geq 0$}.  \nonumber 
\end{align} 

The following theorem states that the lack of coordination in 
deadlock detection initiation by 
fully distributed scheduling will introduce additional overhead in deadlock 
handling. Therefore the fully distributed scheduling in
general cannot perform as efficiently as its centralized counterpart.
 
\begin{theorem} \label{theo:the5}
Let $C_H$ denote the long-run mean average cost under 
fully distributed scheduling with
a random detection interval $Y$ characterized by  
certain distribution $H \in {\cal H}$ with the mean of $\mu$,
and $C(T)$ denote the long-run mean average cost under centralized scheduling 
with a fixed 
detection interval $T$.  Then
\begin{align}
C_H \geq C(T),
\end{align}
when $E(Y)=\mu=T$. 
\hfill  $\blacktriangle$
\end{theorem}

\noindent Proof:
Since the sequence $\{Y_i, i\geq 1 \}$ of 
interarrival times of deadlock detection  
is assumed to be independent of the Poisson deadlock formations,
it is easy to see that the random costs over the intervals 
$(0, Y_1],(Y_1, Y_1+Y_2], \ldots$ are {\it iid}. 
Using the same line of reasoning in  the proof 
of Theorem~\ref{theo:theorem1},   
the long-run mean average cost is expressed as
\begin{align}
C_H =\frac{E(\hbox{random~cost~over~} Y)}{E(Y)},
\end{align}
where $Y \in {\cal H}$ is a random variable representing the interval between two 
consecutive deadlock detections.  
Let $\xi (Y)$ be the random cost in the interval $Y$.
The expected cost over the interval $Y$ is given by 
\begin{align} \label {equ:equ30}
 E ( \xi (Y)) = E\{ E [\xi (Y)|Y]\}  
= \int ^\infty _0 E(C_D+\sum\limits ^{N(y)}_{n=1}C_R(y-S_n)I_{\{N(y)>0\}})dH(y), 
\end{align}
where $S_n=\sum\limits ^n_{i=1}X_i$ 
denotes the time of the {\it n}th deadlock formation and 
$N(y)$ represents the number of independent deadlocks occurred in the time interval $(0,y)$. 
It follows from the independence of  $\{X_i, i\geq 1\}$ and $\{Y_i, i\geq 1\}$, 
and from Eq(\ref{equ:equ30}), 
the long-run mean average cost is 
\begin{align} \label{equ:equ31}
C_H &= \frac{E(\xi (Y))}{E(Y)} =
\frac{\int ^\infty _0 (C_D + \int ^y_0 \lambda C_R(t)dt)dH(y)}{E(Y)} 
= \frac{C_D}{E(Y)} + \frac{\int^\infty _0 \left 
( \int^\infty_t \lambda C_R(t)dH(y) \right)dt}{E(Y)} \nonumber \\
&= \frac{C_D}{E(Y)} + \frac{\lambda \int^\infty_0 
C_R(t) \bar{H}(t) dt}{E(Y)}.
\end{align}

\noindent When $E(Y)=\mu=T$, meaning that 
the fixed deadlock detection interval $T$ equals to 
the mean value of the random detection interval $Y$, 
we compare the centralized (fixed) detection scheduling with the rate of $1/T$ with 
the fully distributed (random) one with the mean rate of $1/E(Y)=1/\mu$.
According to Theorem~\ref{theo:theorem1}, 
the long run mean average cost of fixed detection is given as 
\begin{equation} \label{equ:equ34}
C(T)= \frac{C_D}{\mu} + \frac{\lambda \int ^{\mu}_0 C_R(t)dt}{\mu}.
\end{equation}

\noindent Subtracting Eq(\ref{equ:equ34}) from Eq(\ref{equ:equ31}) yields
\begin{align} \label{equ:equ35}
&C_H-C(T)= \frac{\lambda}{\mu} 
\left \{ \int ^\infty _0 C_R(t)\bar{H}(t)dt - 
\int ^\mu_0 C_R(t)dt \right \} 
= \frac{\lambda}{\mu} \left \{ \int ^\infty _\mu C_R(t)\bar{H}(t)dt -  
\int ^\mu _0 C_R(t)H(t)dt \right \} \nonumber \\ 
&\geq  \frac{\lambda}{\mu} \left \{ C_R(\mu) \int ^\infty _\mu \bar{H}(t)dt - 
C_R(\mu) \int ^\mu _0 H(t) dt\right \}  = \frac {\lambda C_R(\mu)}{\mu} 
\left \{ \int ^\infty _\mu \bar{H}(t)dt -
\int ^\mu _0 (1-\bar{H}(t))dt \right \} \nonumber \\
&=  \frac {\lambda C_R(\mu)}{\mu} \left \{ \int ^\infty _0 
\bar{H}(t)dt - \mu \right \} =0.  
\end{align}
\noindent Hence we have 
\begin{align}\label{equ:tt}
C_H \geq C(T).
\end{align}
Theorem~\ref{theo:the5} is thus established. \hfill $\blacksquare$ 

\noindent It can be seen from Eq(\ref{equ:tt}) 
that $C_H \geq C(T)$ and the equality holds if and only
if $Y$ is a degenerate random variable when
$Prob(Y=T)=1$.
Theorem~\ref{theo:the5} asserts that 
the fully distributed (random) deadlock 
detection scheduling in general results in  
an increased overhead in overall deadlock handling. 

\section{Conclusion}

Deadlock detection scheduling is an important, yet often overlooked aspect  
of distributed deadlock detection and resolution.  
The performance of deadlock handling
not only depends upon per-execution complexity of deadlock detection/resolution algorithms,
but also depends fundamentally upon deadlock detection scheduling and 
the rate of deadlock formation. 
Excessive initiation of deadlock detection results in an increased number of
message exchange in 
the absence of deadlocks, while insufficient initiation of 
deadlock detection incurs an increased cost of 
deadlock resolution in the presence of deadlocks.
As a result, reducing the per-execution cost of 
distributed deadlock detection/resolution algorithms alone does not 
warrant the overall performance improvement on deadlock handling. 

The main thrust of this paper is to bring an awareness to the problem of 
deadlock detection scheduling and its impact on 
the overall performance of deadlock handling.
The key element in our approach is to develop a time-dependent 
model that associates the deadlock 
resolution cost with the deadlock persistence time. 
It assists the study of time-dependent deadlock resolution cost in connection with 
the rate of deadlock formation and the frequency of deadlock detection initiation,
differing significantly from the past research that 
focuses on minimizing per-detection and per-resolution costs.

Our stochastic analysis, which
solidifies the ideas presented 
in \cite{Knapp1987,Singhal1989,Park1992,Krivokapic1999}, 
shows that there exists a unique deadlock detection frequency that guarantees 
a minimum long-run mean average cost for deadlock handling 
when the total number of processes in a distributed system is sufficiently large,
and that the cost of overall deadlock handling grows linearly with
the rate of deadlock formation.

In addition, we study the fully distributed (random) deadlock detection scheduling 
and its impact on the performance of deadlock handling.
We prove that in general the lack of coordination in deadlock detection initiation
among processes will increase  
the overall cost of deadlock handling.  

Theoretical results obtained in this paper 
could help system designers/practitioners to 
better understand the fundamental performance tradeoff between 
deadlock detection and deadlock resolution costs, as well as 
the innate dependency of optimal detection frequency upon
deadlock formation rate.  
However, there are still a lot of questions regarding  
how to use  
theoretical results to fine-tune the performance of a
distributed system.   
Determination of the actual rate of deadlock formation 
and verification of the Poisson process are problems of great 
complexity that can be influenced 
by many known/unknown factors such as the granularity of locking,
actual distribution of resource, process mix, and 
resource request and release patterns \cite{Singhal1989}. 
Tapping into system logging files 
and inferring the actual deadlock formation rate 
via data mining could provide an effective and feasible way 
to translate theoretical insights into actual system performance gain. 

\section{Acknowledgements}
We would like to thank Drs. Marek Rusinkiewicz, Wai Chen and Ritu Chadh at Applied 
Research, Telcordia Technologies for their constructive comments.
We are grateful to the 
anonymous reviewers for critically reviewing the manuscript and 
for their truly helpful comments. 
Yibei Ling would like to especially thank Dr. Shu-Chan Hsu  
in Department of Cell Biology and Neuroscience at Rutgers University 
for her encouragement and support.
\bibliography{deadlock}
\bibliographystyle{plain}
\end{document}